# Magnetic Dynamic Polymers for Modular Assembling and Reconfigurable Morphing Architectures

*Xiao Kuang[§], Shuai Wu[§], Yi Jin, Qiji Ze, S. Macrae Montgomery, Liang Yue, H. Jerry Qi\*, Ruike Zhao\**

Dr. X. Kuang, S. M. Montgomery, Dr. L. Yue, Prof. H. Jerry Qi
The George W. Woodruff School of Mechanical Engineering, Georgia Institute of Technology, Atlanta, GA 30332, USA
E-mail: qih@me.gatech.edu

S. Wu, Y. Jin, Dr. Q. Ze, Prof. R. Zhao
Department of Mechanical and Aerospace Engineering, The Ohio State University, Columbus, OH, 43210, USA
E-mail: zhao.2885@osu.edu

[§] These two authors are equal contribution first authors.

**Keywords:** magnetic soft material, covalent adaptive polymer, modular assembly, shape morphing, reconfigurable architecture

**Abstract**

Shape morphing magnetic soft materials, composed of magnetic particles in a soft polymer matrix, can transform shapes reversibly, remotely, and rapidly, finding diverse applications in actuators, soft robotics, and biomedical devices. To achieve on-demand and sophisticated shape morphing, the manufacturing of structures with complex geometry and magnetization distribution is highly desired. Here, we report a magnetic dynamic polymer composite composed of hard-magnetic microparticles in a dynamic polymer network with thermal-responsive reversible linkages, which permit functionalities including targeted welding, magnetization reprogramming, and structural reconfiguration. These functions not only provide highly desirable structural and material programmability and reprogrammability but also enable the manufacturing of structures with complex geometry and magnetization distribution. The targeted welding is exploited for modular assembling of fundamental building modules with specific logics for complex actuation. The magnetization reprogramming enables altering the morphing mode of the manufactured structures. The shape reconfiguration under magnetic actuation is coupled with network plasticity to remotely transform two-dimensional tessellations into complex three-dimensional architectures, providing a new strategy of manufacturing functional soft architected materials such as three-dimensional kirigami. We anticipate that the reported magnetic dynamic polymer provides a new paradigm for the design and manufacturing of future multifunctional assemblies and reconfigurable morphing architectures and devices.

Shape-morphing materials capable of altering the structural geometry upon external stimuli, such as heat, light, and magnetic field, find diverse applications in actuators,[1, 2] soft robots,[3-5] flexible electronics,[6-8] and biomedical devices.[9-11] Various stimuli-responsive smart materials, including shape memory polymers,[12-15] hydrogel composites,[16-18] liquid crystal elastomers,[19, 20] and magnetic soft materials (MSMs),[21, 22] have been implemented. In particular, MSMs, composed of hard-magnetic materials in a soft polymer matrix, enable remote, fast, and reversible shape morphing, and have attracted increasing attention in applications such as soft robots for minimally invasive surgery,[23-25] where the actuation in confined and enclosed spaces is required. The magnetically actuated shape-morphing is a result of the interactions between the MSM's magnetization (or ferromagnetic polarity) and the applied magnetic field. When the magnetization of embedded hard-magnetic particles is not aligned with the applied magnetic field, a body torque is exerted on the material and leads to a deformation that tends to align the magnetization with the magnetic field.[26-28] The on-demand shape morphing is determined by both the structural geometry and the magnetization distribution. Therefore, manufacturing of structures with intricate geometry and magnetization distribution is highly desirable, which enables shape morphing in a programmable fashion.

From existing efforts, molding with template-assisted post magnetization is very commonly used to fabricate the geometry and magnetization distribution.[29-31] To enhance the material and structural programmability, various advanced manufacturing techniques, including ultraviolet lithography[32, 33] and additive manufacturing,[34-36] have recently been developed to create complex shapes and magnetization distribution via physical alignment of magnetic dipoles followed by chemical curing of the matrix. For the above methods, the magnetization is coupled with the manufacturing process, retarding the manipulation of the geometry and the magnetization distribution after manufacturing. To tackle this issue, magnetic-assisted material assembling and magnetization reprogramming approaches have been recently explored. In the magnetic-assisted material assembling, the intrinsic dipole-dipole interactions enable the self-organization of magnetic modules into larger two-dimensional (2D) structures.[37-40] Modules with hard magnets as inclusions are often used to enhance the magnetic attraction force, however, it compromises the structure's overall flexibility and homogeneity, which is not preferred if magnetic actuation with complex shape morphing is needed.[37, 38] In general, most existing magnetic-assisted assemblies are based on physically attaching the magnetic modules to each other, which can be easily broken under external perturbation. Directly reprogramming magnetization distribution provides an alternative approach to post-manipulate actuation modes.[41-45] In this case, remagnetizing the composite using a large magnetic field above the coercivity of the magnetic material,[41, 42] or heating the magnetic material to above its Curie temperature to demagnetize before re-magnetizing it [43, 44] are the most widely used methods. Another recently reported effort reprograms the magnetization by physically realigning the magnetic particles confined in fusible polymer microspheres that are embedded in a deformed elastomer matrix.[45] Although reprogramming magnetization provides reconfigurable shape morphing, the material's function can still be limited by its unchangeable structural geometry.

In this work, we develop a magnetic dynamic polymer (MDP) that enables both structural and material programmability and reprogrammability with complex geometry and magnetization distribution for multifunctional and reconfigurable shape morphing. The MDP consists of hard-magnetic microparticles (NdFeB) in a thermally reversible cross-linked dynamic polymer (DP) matrix (**Figure 1a**). As a proof of concept, a DP matrix bearing thermally reversible Diels-Alder reaction between furan and maleimide is used. The maleimide and furan groups can proceed with forward Diels-Alder (DA) reaction at low temperature to form adduct linkages, which can be cleaved via retro Diels-Alder (rDA) reaction upon heating.[46, 47] The dynamically cross-linked DP and MDP can rearrange network topology by bond exchange reaction (BER) at mild temperature ($T_{BER}$) and go through bond cleavage at elevated temperature ($T_{rDA}$), enabling the attractive property of heat-induced reversible elastic-plastic transition after material manufacturing (**Figure 1b**). The MDP integrates three functional properties, including (i) welding-enhanced assembling, (ii) in-situ magnetization reprogramming, and (iii) remotely controlled permanent shape reconfiguration. We first illustrate the seamless welding of modules via forming new dynamic linkages at the interfaces to achieve modular assembling (**Figure 1c**). With the cleavage of linkages at $T_{rDA}$, a small magnetic field can chain up dipoles along the field direction for magnetization reprogramming, which further alter the shape morphing modes under the actuation magnetic field (**Figure 1d**). Moreover, simultaneous magnetic actuation and structural reconfiguration of the DP matrix at mild temperatures ($T_{BER}$) allow remotely reshaping the material into a new stress-free architecture (**Figure 1e**). Besides, the MDP based materials and architectures with programmed shape and magnetization distributions are capable of fast and reversible morphing shapes under external magnetic fields at room temperature.

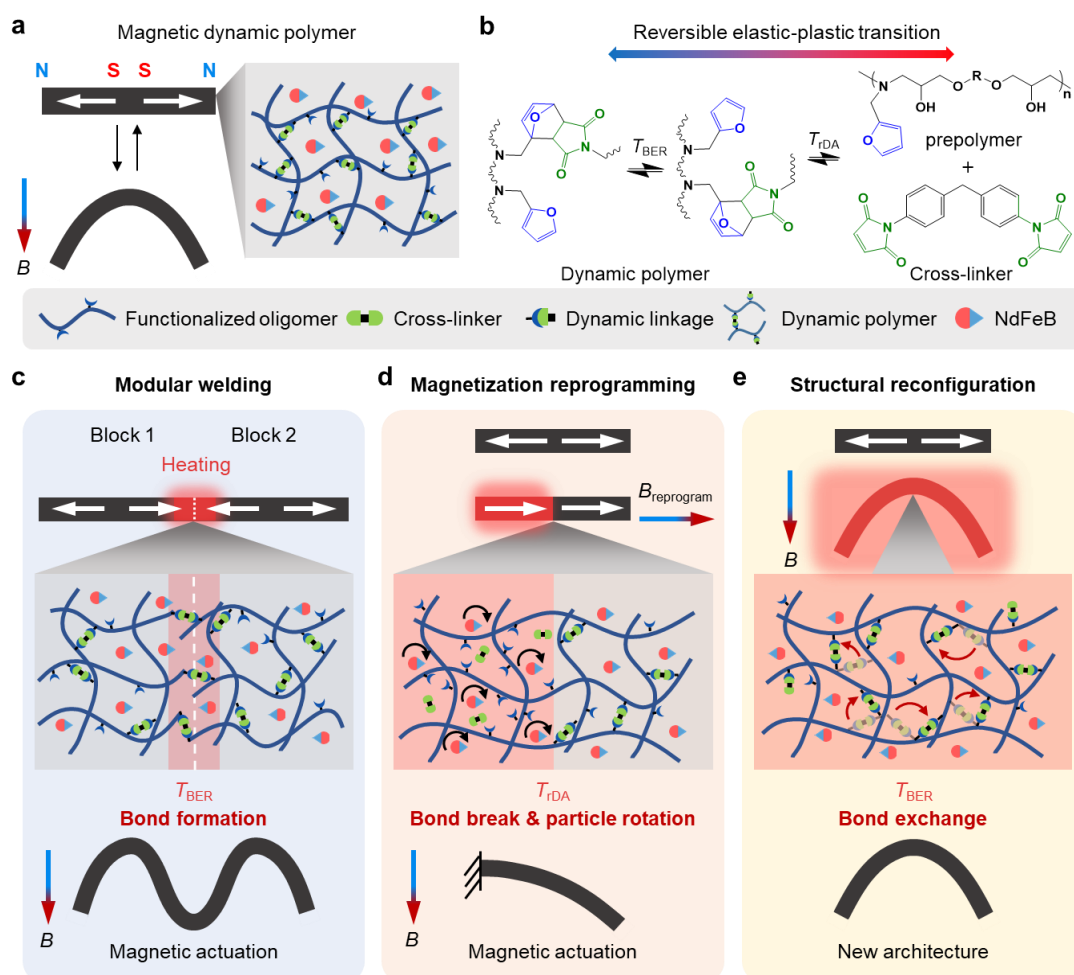

**Figure 1.** Schematics of the working mechanism and functions of the magnetic dynamic polymer (MDP). a) Schematics of the MDP composition. NdFeB microparticles are embedded in a dynamic polymer (DP) bearing reversible chemical bonds. b) Scheme of reversible elastic-plastic transition via network topology transition in the Diels-Alder reaction-based DP at different temperatures. The bond exchange reaction between free furan and Diels-Alder adduct linkage for network arrangement is predominant at mild temperatures ($T_{BER}$), and reversible bond cleavage is favored at elevated temperature ($T_{rDA}$). c) Schematics of modular assembly and seamless welding of MDP modules at a temperature near $T_{BER}$. d) Schematics of magnetization reprogramming by bond cleavage and dipole rotation under a magnetic field at $T_{rDA}$. e) Schematics of the magnetically guided structural reconfiguration of MDP by plasticity via stress relaxation at $T_{BER}$.

To prepare the DP matrix, a furan grafted prepolymer is first synthesized by the ring-opening reaction between an epoxy oligomer and fururylamine. The prepolymer chains have an average of 13 pending furan groups, as indicated by the gel permeation chromatography (GPC) measurement (Figure S1). The linear prepolymer is cross-linked by the bismaleimide cross-linker via DA reaction, forming thermally reversible adduct linkages, as evidenced by Fourier transform infrared (FTIR) spectroscopy (Figure S2). The maleimide to furan ratio ($r$) is used to tune the mechanical property, network cross-linking density, and thermal property of DP (Figure S3). The DP with $r$ = 0.15 (Young's modulus of 106 ± 9 kPa and glass transition temperature of -35 ºC) is selected for use in the rest of this paper (**Figure 2a**). To prepare MDP, NdFeB particles with an average size of 25 μm are dispersed in the DP matrix (Figure S4). The DP and MDP are soft stretahcale elastomers showing break strain over 200%. Young's modulus of MDP increases linearly from 106 kPa to 515 kPa with NdFeB microparticles concentration varying from 0 to 20 vol%, respectively (Figure S5). We choose the NdFeB particle concentration of 15 vol% to manufacture MDP with low stiffness and ideal magnetic properties for efficient magnetic actuation. After manufacturing and post magnetization, the obtained MDP shows Young's modulus of 400 ± 20 kPa and magnetization of 75 kA m$^{-1}$ (Figure S6).

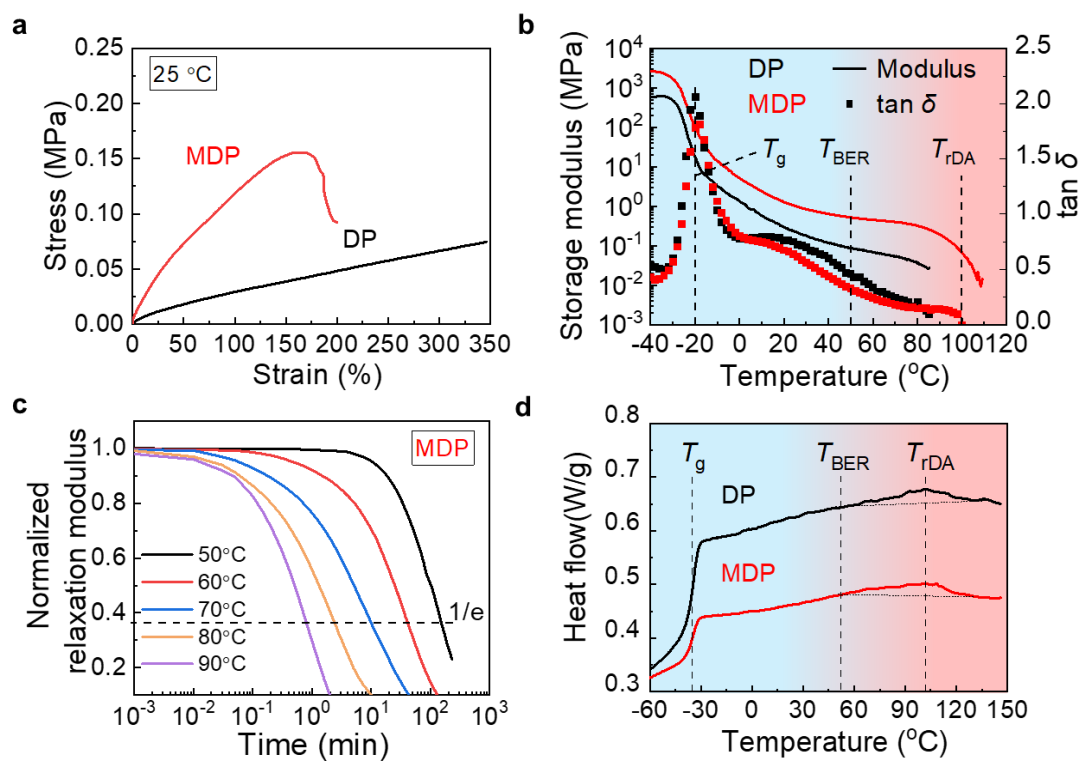

**Figure 2.** Mechanical and thermomechanical characterization of the DP and MDP. a) Tensile stress-strain curves of the DP and MDP. b) DMA heating curves for the DP and MDP from -40 to 120 °C with the marked characteristic glass transition temperature ($T_g$), bond exchange reaction temperature ($T_{BER}$), and retro Diels-Alder reaction temperature ($T_{rDA}$). c) Normalized relaxation modulus as a function of time in the stress relaxation test of MDP at mild temperatures from 50 to 90 °C. d) DSC heating curves of DP and MDP from -60 to 150°C with marked characteristic temperatures.

Unlike conventional chemically cross-linked polymer and composites, DP network or covalent adaptable network polymer is a class of chemically cross-linked polymer with a sufficient amount of dynamic or exchangeable covalent bonds in the polymer network, which enables material flow and permanent shape change after activating the dynamic bonds.[48-52] DP networks possess desirable attributes of chemical cross-linking in thermosets and reprocessability in thermoplastics.[53-56] The fabricated DP and MDP here manifest heat-induced reversible elastic-plastic transition. At room temperature, DP and MDP show excellent elasticity and low hysteresis in the cyclic loading-unloading test (Figure S7), due to low $T_g$ and stable chemical linkages. This is also evidenced by a rubbery plateau from 25 to 80 °C by dynamic mechanical analysis (DMA) test (**Figure 2b**). The rubbery modulus of the MDP slightly decreases before 80 °C and sharply drops after 100 °C, owing to the cleavage of the dynamic linkages. Note that the MDP shows a larger high-temperature modulus and network stability than the DP due to the potential polymer-filler interaction. Furthermore, creep tests are conducted at different temperatures and applied stress to reveal temperature-sensitive mechanical properties of the MDP. The apparent zero shear rate viscosity ($\eta_0$) of the chemically cross-linked network is obtained by the creep test (Figure S8). The $\eta_0$ of the MDP decreases from nearly $10^9$ Pa·s at room temperature to $10^6$ Pa·s at 90 °C, showing prominent temperature-sensitive mechanical behavior. Aside from the temperature-dependent mechanical properties, the network topology can also be rearranged via dynamic bond exchange reaction,[57] which is evaluated by the stress relaxation tests at various temperatures from 50 to 90 °C. **Figure 2c** shows the normalized relaxation modulus decreasing with time. The relaxation process is accelerated at higher temperatures leading to prominent plasticity. According to the Maxwell model, the network relaxation time ($\tau_{DP}$), defined by the time needed to decay to 36.7 % of original modulus, decreases from 153 min (at 50 °C) to 46 s (at 90 °C). The underlying mechanism for temperature-sensitive elastic-plastic transition upon heating is the network breaking and rearrangement via dynamic bond cleavage and exchange, respectively. To evaluate the temperature-variant bond breaking, differential scanning calorimetry (DSC) testing is conducted. On the DSC heating curves, an endothermal peak ranging from 50 °C to 120 °C is observed in both the DP and the MDP due to continuous bond-breaking via rDA reaction (**Figure 2d**). The rDA reaction degree as a function of temperature is semi-quantitatively evaluated by the normalized areal integration under the endothermal peak of the DSC heating curve (Figure S9). A gel point conversion ($p_{gel}$~74%) is predicted between 80 °C-90 °C (see Supporting Information, dynamic polymer relaxation analysis). The peak temperature at around 100 °C suggests the fastest bond-breaking rate, which is denoted as rDA temperature ($T_{rDA}$). It is noted that the cleaved linkages can reform upon cooling and annealing as supported by FTIR (Figure S2c). On the basis of the reversible elastic-plastic transition of the MDP, $\tau_{DP}$ and magnetic dipole relaxation time ($\tau_{diople}$) (see Supporting Information, magnetic dipole relaxation analysis) can be tuned to manipulate the shape and magnetization by corporative control of heating temperature and magnetic field for multifunctional reconfigurable morphing architectures.

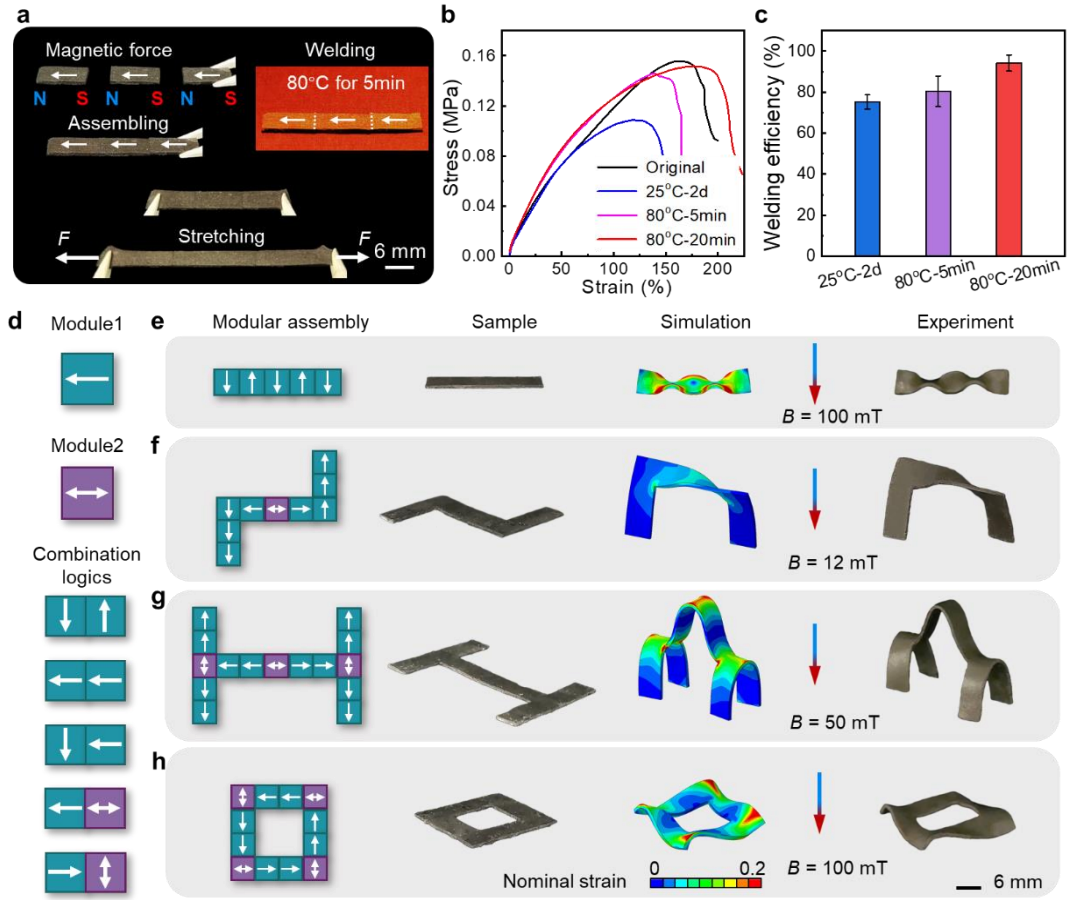

**Figure 3.** Magnetic-assisted modular assembling with seamless welding. a) Images of a long strip assembly consisting of three MDP modules via magnetic attraction followed by near-infrared light (NIR) heating (80 °C for 5 min). b) Tensile stress-strain curves of the original sample and the welded MDP processed samples under different conditions. c) Effect of processing conditions on the welding efficiency of the processed MDP. d) Schematics of a square single-directional magnetization module and a bidirectional magnetization module with five double-unit combination logics via magnetic attraction. e-h) Schematic designs, finite-element simulations, and experimental results of various assembled 2D planar structures with programmed magnetization for complex shape morphing: a twisting strip (e), a 'Z'-shape structure (f), an 'H'-shape structure (g), and a square annulus structure (h).

We first demonstrate the function of magnetic-assisted assembling by magnetic attraction and seamless welding of magnetic modules. **Figure 3a** shows the longitudinally magnetized rectangular magnetic modules automatically attaching to each other rapidly (~ 0.3 s) with good contact. Afterward, the assembly is treated at 80 ºC for 5 min either by direct heating or near-infrared (NIR) light (Figure S10) to be welded into an intact part. As the bond exchange reaction between the contacting interfaces generates strong chemical bonding, the welded assembly can sustain large stretching without breaking (Video S1, Supporting Information). The welding efficiency is quantified by the tensile fracture strain ratio of the welded samples to the original one (**Figure 3b**). The welding efficiency is 75% at room temperature after two days due to the very slow dynamic reaction at room temperature (**Figure 3c**). When at 80 ºC, the welding efficiency can increase to 82 % for 5 min treating and 95 % for 20 min.

We further extend this concept to the modular assembly of stable 2D structures with on-demand magnetization patterns and geometries using different modules. We achieve this by using two basic simple building modules: a square with single-directional magnetization (*Module 1*) and a square with bidirectional magnetization (*Module 2*) (**Figure 3d**). The magnetic attraction provides these two basic building modules with five different combination logics (Video S2, Supporting Information). Under certain boundary conditions, the response of these combination logics under an out-of-plane magnetic field generates twisting, bending, twisting-bending, bending-folding (same direction), and bending-folding (orthogonal direction), respectively. These logics can be further used to achieve more intricate shape changes. For example, a strip with five pieces of *Module 1* arranged in twisting logics is assembled and welded at 80 ºC for 20 min. As shown in **Figure 3e**, the obtained straight strip assembly transforms to a twisting shape rapidly upon applying a magnetic field of 100 mT and quickly recovers its original shape upon the removal of the applied field (Video S2, Supporting Information). The shape morphing of the assembled MDP is predicted by finite-element analysis (FEA) through a user-defined element subroutine,[26] which guides the design of assemblies for target shape transformation under the application of magnetic fields. The simulation is conducted using the experimentally measured mechanical and magnetic properties, showing good agreement with the experimental result.

The logics presented above are further used to assemble the building modules into a rich set of more intricate 2D magnetization patterns on planar structures for complex shape morphing (Video S2, Supporting Information). In **Figure 3f**, the combination of bending, bending-folding (same direction), and bending-twisting logics is used to design a 'Z'-shape structure for large twisting morphing under an applied magnetic field of 12 mT, as predicted by our FEA simulation. An 'H'-shape structure is assembled by using fifteen modules with only bending and bending-folding (both types) logics to generate large pop-up deformation under an external magnetic field of 50 mT as illustrated in **Figure 3g**. We also assemble a closed shape by using bending and bending-folding (both types), displaying an undulating shape morphing under a magnetic field of 100 mT (**Figure 3h**). The simulation results match the experimental results well, demonstrating that the modular assembling strategy can be assisted by FEA to design the shape-morphing architectures and their response under magnetic actuation. It is noted that the magnetization of each module is retained after welding, which is attributed to much longer $\tau_{diople}$ than $\tau_{DP}$ at $T_{BER}$ ($\tau_{diople} \gg \tau_{DP} > 0$) (see Supporting

Information, magnetic dipole relaxation analysis). Our method of using the MDP for magnetic-assisted modular assembling with welding enables material integrity for complicated shape morphing. Moreover, our assembling strategy of using the logics can achieve an almost unlimited number of possible shape morphing designs.

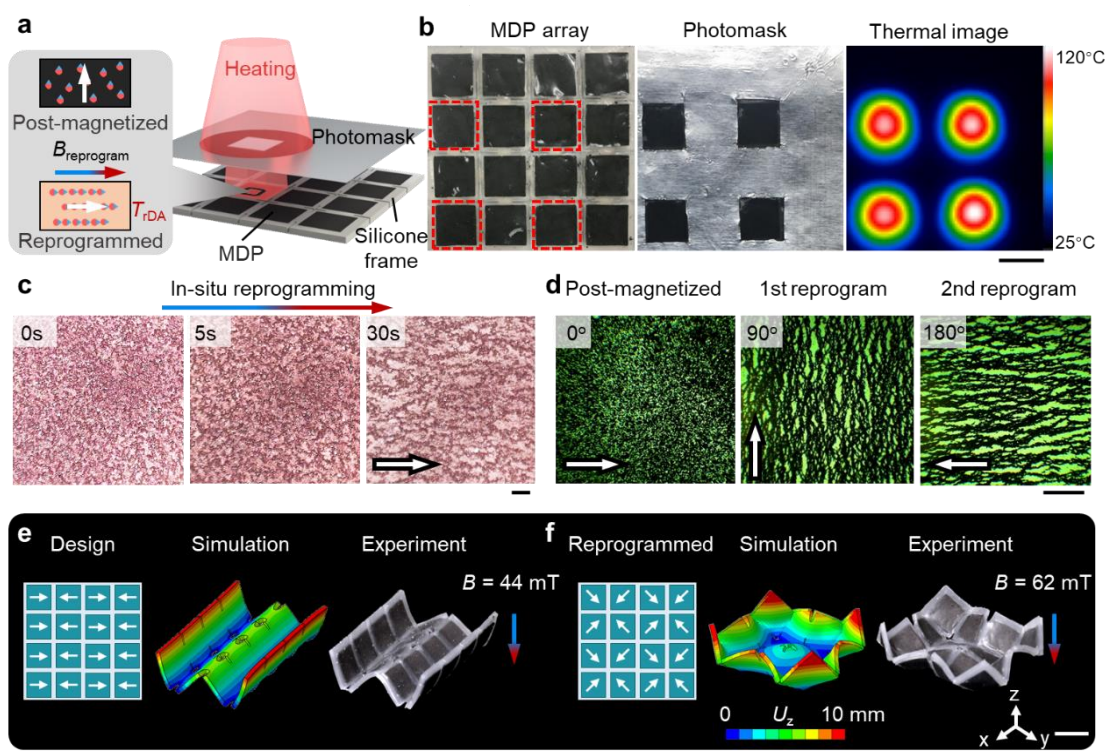

**Figure 4.** In-situ reprogramming of magnetization in the MDP. a) Schematics of magnetization reprogramming of silicone encapsulated MDP array using an NIR light source and a photomask under a magnetic field. The magnetic particles can be realigned by an external magnetic field at $T_{rDA}$. b) Image of an as-fabricated 4 × 4 MDP array, an aluminum photomask, and temperature profile after exposure to NIR illumination. c) Microscopic snapshots of the in-situ realignment of NdFeB particles in MDP at around 110 ºC under a 35 mT magnetic field. d) Multiple cycles of magnetization reprogramming from the initial isotropic dispersion to chain-like structures along the external magnetic field direction. e-f) Designs, simulations, and experimental results of the MDP array with the initial magnetization (e), and reprogrammed magnetization (f). Scale bars in c-d, 500 μm. Scale bars in b, e, f, 10 mm.

We also achieve in-situ magnetization reprogramming via cleavage of dynamic linkages at elevated temperatures. To selectively reprogram the magnetization, photothermal heating by NIR light illumination and photomasks is used (**Figure 4a**). Upon heating above $T_{rDA}$, the viscosity is reduced tremendously, and thus the magnetic particles can move freely in the MDP, leading to dipole realignment along a small external magnetic field, which can be relocked after cooling. As an example, in **Figure 4b**, we fabricate a 4 × 4 MDP array encapsulated by silicone rubber to illustrate the selective magnetization reprogramming (Figure S11 and Video S3, Supporting Information). An aluminum foil as a photomask is used to enable local photothermal heating of selectively exposed MDP cells to 110 - 120 °C. The magnetic particles in the heated MDP cells physically rotate and reorient under a 35 mT homogenous magnetic field provided by a Halbach array magnet. The in-situ dipole realignment is observed under digital microscopy coupled with a pair of electromagnetic coils (Figure S12). As shown in **Figure 4c**, when the MDP viscosity reduces upon heating, the post-magnetized NdFeB particles in the form of small aggregation can rotate within 5 s and then tend to form a chain-like structure after 30 s under a 35 mT magnetic field (Video S3, Supporting Information). Besides, due to the reversible elastic-plastic transition of MDP, the magnetization realignment can be repeated along external magnetic fields with different directions (**Figure 4d**). To test the efficacy of magnetization reprogramming, the planar 4 × 4 MDP array is initially magnetized horizontally in an alternating manner for each column of cells, as shown in **Figure 4e**. The 2D array morphs into a three-dimensional (3D) 'W' shape in a 44 mT out-of-plane magnetic field, which is well captured by the FEA simulation. Next, a four-step reprogramming procedure is used to reprogram the selective MDP cells with four different reprogrammed magnetization directions, respectively (Video S3, Supporting Information). As shown in **Figure 4f**, the targeted reprogrammed magnetization has the polarity oriented in the diagonal of each cell pointing to the four centers. Upon applying a 62 mT out-of-plane magnetic field, the reprogrammed MDP array morphs into a 3D surface with four dents. The experimental result matches the simulation well, suggesting retained magnetization after reprogramming. It is noted that the thermally reversible reaction also enables hot recycling of MDP (by both solvent and hot-compression) (Figure S13). Compared with the existing magnetization reprogramming techniques for various magnetic soft composites, our MDP offers additional advantages of selective magnetization reprogramming under mild conditions, large remanent magnetization for efficient actuation, and sustainability.

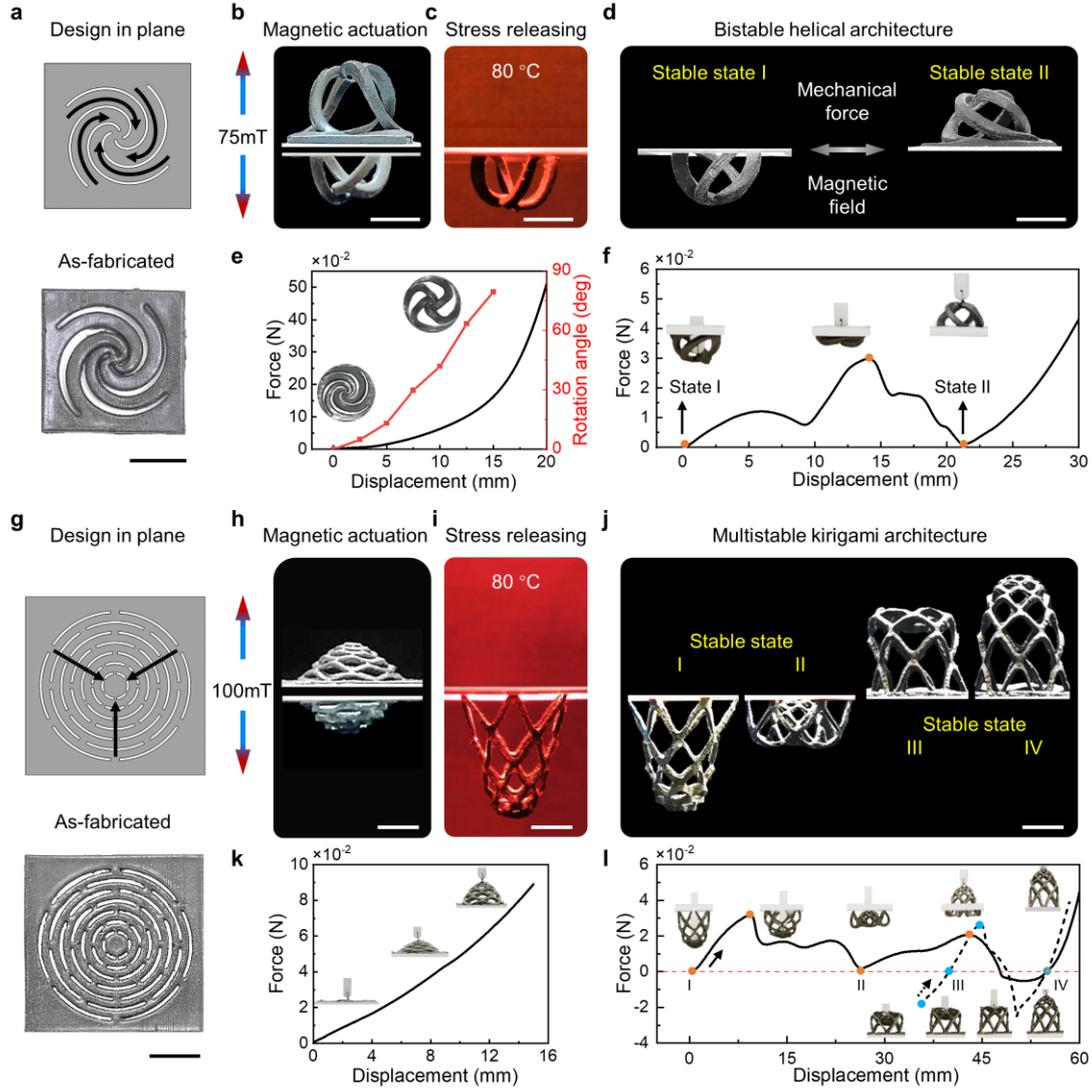

**Figure 5.** Magnetic-assisted 3D structural reconfiguration of the MDP for complicated multistable architectures. a) Schematic design and the as-fabricated planar kirigami with helical cuts (helical design). Black arrows indicate magnetization. b) Magnetic actuation of the planar helical structure. c) Stress releasing process of the magnetic-actuated helical structure under the NIR irradiation. d) Mechanical or magnetic actuation of reshaped bistable helical architecture. e-f) The force/rotation angle-displacement curves and snapshots of the helical design in tensile test: initial planar structure (e) and reshaped 3D architecture (f). g) Schematic design and the as-fabricated planar kirigami with concentric arc cuts (concentric arc design). h) Magnetic actuation of the planar concentric arc structure. i) Stress releasing process of the magnetic-actuated concentric arc structure under the NIR irradiation. j) Images of the multistable 3D kirigami architecture with four stable states. k-l) The force-displacement curves and snapshots of the concentric arc design in tensile test: initial planar structure (k) and reshaped 3D architecture (l). Scale bars for all images, 10 mm.

The reshaping capability offered by the plasticity of the MDP can be harnessed to manufacture 3D free-standing intricate architectures with complex magnetization that would otherwise be very challenging to implement. In **Figure 5a,** a planar kirigami design with four helical cuts (or helical design) is first molded and then post-magnetized in its mechanically stretched-out configuration (Figure S14). The fast and reversible magnetic actuation into a helical shape under a 75 mT out-of-plane magnetic field is shown in **Figure 5b** (Video S4, Supporting Information). When applying a magnetic field and NIR heating (~80 °C) simultaneously, actuated helical shape gradually releases its internal stress due to the bond exchange in the polymer network (**Figure 5c**). Upon removing the magnetic field and the heating source after ~15 min, a stress-free 3D helical architecture with self-adjusted magnetization is formed. The reshaped helical architecture also comes with bistability, which can be triggered either by magnetic actuation or mechanical loading (**Figure 5d** and Video S4, Supporting Information). Such a 3D architecture would be very challenging to manufacture but can be easily achieved using our MDP approach. We conducted tensile tests to compare the mechanical behaviors and properties of the as-fabricated 2D design and the reshaped 3D architecture. Both structures are loaded by stretching the central point while fixing the edges, with results illustrated in **Figure 5e, f** (also see Figure S14 and Video S5, Supporting Information). Before the reshaping process, the 2D structure is monostable, as the force (black curve) and rotation angle (red curve) is monotonically increasing with the displacement (**Figure 5e**). After reshaping, the 3D helical architecture exhibits bistable behavior, which is precisely characterized by the force-displacement curve from the tensile test (**Figure 5f**) loaded from the initial downward pop-out stable state.

During loading, the architecture reaches the critical snapping point at around 14 mm displacement, and the structure quickly snaps through to the upward pop-out state at 21 mm displacement, where the force reduces to zero.

A more complicated 3D kirigami architecture based on concentric arc design with multistable states is manufactured by the same MDP stress releasing approach. **Figure 5g** shows the design and the as-fabricated planar kirigami geometry comprising seven layers of concentric arcs connected by hinges, which is initially magnetized in a mechanically stretched-out configuration (Figure S15). Its magnetic actuation under a 100 mT out-of-plane magnetic field is shown in **Figure 5h**. After exposing to the NIR light and an out-of-plane magnetic field for about 30 min (**Figure 5i**), the planar concentric arc structure transforms into a stress-free kirigami architecture with four stable states, as illustrated in **Figure 5j** (see Video S6, Supporting Information). The different mechanical behaviors before and after the reshaping process are evaluated by tensile tests with the same boundary conditions as previous (Figure S15). **Figure 5k** shows that the initial planar kirigami is monostable. In contrast, as illustrated in **Figure 5l**, the reshaped 3D kirigami architecture has four stable states and three snapping points (marked by orange and blue circles for two loading paths) (see Figure S15 and Video S7, Supporting Information). By either magnetic actuation or mechanical stretching starting from the *stable state I*, two stable states *II* and *IV* are captured with two snapping points identified between *State I* and *II*, and *State II* and *IV*, respectively. This loading path is illustrated by the black solid curve and top-row insets. Note that after the architecture snaps up at the second snapping point, it compresses the load cell until reaching *stable state IV*. *Stable state III* can be obtained from a different loading path when stretching from an intermediate compressed state, indicated by the black dashed curve and bottom-row insets of **Figure 5l**. This state shows an overall upward configuration with the middle part pushed down. After reaching the critical high-energy point, the middle part snaps up, and the architecture achieves *stable state IV* (Video 7, Supporting Information). To our best knowledge, this is the first time that complex 3D architectures are manufactured through simple remotely controlled shape morphing of planar structures. The stress releasing of the MDP can be a versatile approach to manufacture functional free-standing 3D architectures such as 3D kirigami and origami tessellations, which finds broad applications in morphing architectures and metamaterials with programmable complex geometries and novel mechanical properties.[58-61]

The MDP's merits are emphasized through its covalent adaptive network and magnetic-responsive feature. Comparing with previous shape programmable materials (Figure S16 and Table S1), MDP stands out with outstanding performance including untethered, fast, and reversible actuation, as well as excellent material and structural programmability and reprogrammability. Besides, MDPs show prominent functional properties, including welding, reshaping and recycling. Thus, the multifunctional reconfigurable morphing behaviors distinguish MDPs from the existing shape morphing and other magnetic soft materials[21, 62].

In summary, we report a magnetic dynamic polymer composite to create structures with complex geometry and magnetization distribution for modular assembling and reconfigurable shape morphing architectures. The dynamic polymer network rearrangement and the magnetic dipole relaxation are tuned by cooperative control of the temperature field and the magnetic field. Functional properties and applications, including seamless welding of modular assembly with targeted functional actuation, magnetization reprogramming for reconfigurable actuation modes, and remote-controlled structural reconfiguration with unusual properties, are demonstrated. The concept of the magnetic dynamic polymer by merging covalent adaptive network polymers and magnetic materials can be extended to diverse magnetic soft materials, using various stimuli-responsive dynamic reactions and numerous magnetic materials, with tunable mechanical, rheological, and magnetic properties. As magnetic soft materials gain increasing attention, the magnetic dynamic polymer can influence many areas beyond reconfigurable shape morphing. The unique performance of welding and remote-controlled structural reconfiguration in manufacturing, healing during service, and recycling at the end of service can provide a green material and enhance functionality of magnetic soft materials. We envision the magnetic dynamic polymer and its derived functions offer great potentials for next-generation multifunctional assemblies, reconfigurable shape morphing architectures and devices.

**Experimental Section**

*Furan grafted prepolymer synthesis*: 50 g Poly(ethylene glycol) diglycidyl ether and 9.8 g furfurylamine in the stoichiometric ratio were mixed in 15 g dimethylformamide (DMF)，and 0.374 g of 2, 6-Di-tert-butyl-4-methylphenol was dissolved to the above solution in a round bottle flask. After degassing under vacuum for 5 min, the mixture was sealed for the reaction by stirring at 80 °C for 20 h and then 110 °C for another 4 h before cooling down. The obtained viscous dark yellow solution was stored at room temperature before use. For prepolymer characterization, the solution was dried in a vacuum oven at 70 °C for over one day. All reagents were purchased from Sigma-Aldrich (St. Louis, MO, USA) without further purification.

*Dynamic polymer and magnetic dynamic polymer preparation*: Dynamic polymer (DP) was prepared by cross-linking the furan grafted prepolymer with bismaleimide cross-linker (Sigma-Aldrich, St. Louis, MO, USA). Briefly, linear prepolymer solution (containing 20 wt% of solvent) and bismaleimide with various maleimide/furan ratios (*r*) were manually mixed at room temperature. To manufacture magnetic dynamic polymer (MDP), 15 vol% NdFeB microparticles (134.0 wt% to the polymer matrix) with an average size of 25 μm (Magnequench, Singapore) were added into the prepolymer and cross-linker mixture (with $r = 0.15$) followed by manually mixing to obtain a homogeneous blend. The resin mixture and the composite resin were poured on a polytetrafluoroethylene (Teflon®PTFE, McMaster-Carr, Elmhurst, IL, USA) film. The curing was conducted with precuring at 50 °C for around 60 min to form a soft gel and then post-treated in a vacuum oven at 60 °C for one day to remove most of the solvent. To obtain films with controlled thickness, the samples were hot compressed at 110 °C with spacers and PTFE film separator. The samples were cooled down and stored at room temperature for at least one day before characterization. The materials were post-magnetized with an initial configuration under impulse magnetic fields (1.5 T) generated by an in-house built impulse magnetizer.

*Characterization.* The uniaxial tensile tests, dynamic thermomechanical measurements, and stress relaxation tests were performed on a dynamic mechanical analysis (DMA) tester (DMA 800, TA Instruments, New Castle, DE, USA) using rectangular samples (dimension: about 25 × 4 × 0.9 mm). In the uniaxial tension tests, the strain rate was 0.5 min$^{-1}$ and three specimens were tested for each type of the samples for reporting the average results. The Young's modulus was calculated by the secant modulus at 0.5% strain. DMA test was conducted using a constant oscillation amplitude of 30 μm, a frequency of 1 Hz, and a force track of 125 %. The temperature was ramped from -60 ºC to 120 ºC at a heating rate of 3 ºC min$^{-1}$. In the stress-relaxation experiments, the samples were equilibrated at a predetermined temperature for 10 min, and then a constant strain of 2 % was applied to monitor the evolution of stress as a function of time. Differential Scanning Calorimetry Testing (DSC) was measured on a Q200 DSC (TA Instruments, New Castle, DE, USA) using T-zero aluminum pans under a nitrogen purge. The testing temperature ranges from −80 to 150 °C with a heating and cooling rate of 5 ºC min$^{-1}$. The temperature field of samples during heating was captured using a Seek Thermal CompactPRO thermal image camera (Tyrian Systems, Inc., Santa Barbara, CA, USA)

*Finite-element analysis.* The shape actuation of MDP architectures and array under external magnetic fields were simulated using a user-defined element subroutine implemented in the finite-element analysis software ABAQUS 2019 (Dassault System, Dassault System, Providence, RI, USA). The input parameters were used: Young's modulus $E = 400$ kPa for MDP and $E=130$ kPa for silicone rubber, the bulk modulus $K = 1,000\, E$ (approximate incompressibility), the magnetization of the MDP $M_r = 75$ kA m$^{-1}$ (15 vol% of magnetic particles), and the uniform external magnetic field.

**Acknowledgments**

X. K. and S. W. contribute equally to this work. R.Z., Q.Z., S.W., and Y.J. acknowledge support from NSF Career Award CMMI-1943070 and NSF Award CMMI-1939543. H.J.Q., X.K, S.M and L.Y. acknowledge the support of AFOSR grants (FA9550-16-1-0169 and FA-20-1-0306; Dr. B.-L. "Les" Lee, Program Manager), the gift funds from HP, Inc. and Northrop Grumman Corporation. This work was performed in part at the Georgia Tech Institute for Electronics and Nanotechnology, a member of the National Nanotechnology Coordinated Infrastructure, which is supported by the National Science Foundation (ECCS-1542174).

**Conflict of Interest**

The authors declare no conflict of interest